\documentclass[11pt]{article}
\usepackage{amsmath,amssymb}
\usepackage{cite}
\usepackage{graphicx}
\title{The Higgs Field and the Jordan Brans Dicke Cosmology}
\author{Onder Dunya\footnote{onder.dunya@boun.edu.tr}, \ Levent Akant\footnote{levent.akant@boun.edu.tr}, \ Metin Arik\footnote{metin.arik@boun.edu.tr}, \ Yelda Kardas\footnote{yelda.kardas@boun.edu.tr}, \\ 
Selale Sahin\footnote{selale.sahin@boun.edu.tr} \ and \ Tarik Tok\footnote{tarik.tok@boun.edu.tr} \\
\textit{Department of Physics, Bogazici University, Bebek, Istanbul, Turkey}}
\begin{document}
\maketitle

\begin{abstract}

We investigate a field theoretical approach to the Jordan-Brans- Dicke (JBD) theory extended with a particular potential term on a cosmological background by starting with the motivation that the Higgs field and the scale factor of the universe are related. Based on this relation, it is possible to come up with mathematically equivalent but two different interpretations. From one point of view while the universe is static, the masses of the elementary particles change with time. The other one, which we stick with throughout the manuscript, is that while the universe is expanding, particle masses are constant. Thus, a coupled Lagrangian density of the JBD field and the scale factor (the Higgs field), which exhibit a massive particle and a linearly expanding space in zeroth order respectively, is obtained. By performing a coordinate transformation in the field space for the reduced JBD action whose kinetic part is nonlinear sigma model, the Lagrangian of two scalar fields can be written as uncoupled for the Higgs mechanism. After this transformation, as a result of spontaneous symmetry breaking, the time dependent vacuum expectation value (vev) of the Higgs field and the Higgs bosons which are the particles corresponding to quantized oscillation modes about the vacuum, are found.

\end{abstract}

\maketitle

\section{Introduction}
The concept of mass has been very important and challenging to understand at the fundamental level throughout the advancement of modern physics. The particle physics as well as the foundations of classical physics such as the dynamics of particle motion and the phenomenon of gravitation, in fact depend on this concept. Lately by the discovery of the Higgs boson\cite{2012observation}, one of the most striking developments in our understanding of this concept has been the fact that the mass in physics is fundamentally resulting from the vacuum expectation value of the Higgs field\cite{higgs1964}. Also, all the elementary particles in the standard model obtain their masses by this mechanism\cite{weinberg1991}\cite{salam1980}\cite{glashow1980} which is only consistent in Minkowski spacetime. Although from a historical perspective the concept is divided into two as inertial and gravitational masses, their equivalence which is known the weak equivalence principle (WEP), is essential by experiments\cite{eotvos1922}. WEP ensures test bodies follow the same path in a gravitational field regardless of their compositions. So, this equivalence was one of the main motivations for Einstein to construct the general theory of relativity which explains the gravitation in a purely geometrical way. Another motivation was Mach's principle which relates an inertial force on a body to the gravitational effects originating from the matter distribution of the universe. While Newton's concept of absolute space defines a special frame of reference and an inertial force is the result of motion relative to this frame, Mach's principle states that the observable motion is the relative one and there is no a special frame of reference. Thus, based upon Mach's principle, a test particle experiences an inertial force because of its relative motion to the rest of the universe, or simply, the physical space shaped by distant stars and galaxies. Furthermore, a model which relates inertia to the gravitational potential of the universe, has been proposed by Sciama\cite{sciama1953}. In the rest of this section, a novel connection between inertial mass and the metric tensor is constructed by means of the Higgs field. A similar model\cite{arik2020} has recently been introduced by two of the authors and the more complete and the precise one in terms of theoretical arguments and calculations is studied in this manuscript.

In the general theory of relativity, motion of particles are determined by the action
\begin{equation}
S=-m\int ds=-m\int \sqrt{g_{\mu\nu}dx^\mu dx^\nu},
\end{equation}
where $m$ and $ds$ are the mass of the particle and the length of the Riemannian line element respectively. As is stated earlier, since, from a field theoretical point of view, a particle is gained mass because of its interaction with Higgs field $\phi$, the mass of the particle should be time dependent if the vev of the Higgs field changes with time throughout the evolution of the universe. However, the variation of the field must be at a sufficiently slow rate so that the concept of mass is not put under too much stress and the factor in front of the interaction term is interpreted as mass in a field theoretical Lagrangian density. So, once the solution of the Higgs field is obtained in terms of the parameters of our model, it will be shown that this condition is satisfied. Based on this motivation, the same action in Eq.(1) can be written as
\begin{equation}
S=-m_0\int \frac{\phi(t)}{\phi_0} \sqrt{g_{\mu\nu}dx^\mu dx^\nu}.
\end{equation}

Another fundamental understanding of the universe is the concept of the expanding universe as described by the FLRW metric tensor\cite{friedmann1922,lemaitre1931,robertson1935,walker1937} for which the line element is given by 
\begin{equation}
ds^2=a^2(t)(-dt^2+dx^2+dy^2+dz^2)
\end{equation}
where t is, in cosmological language, called conformal time. Here and henceforth we will use units $\hbar=c=1$. If the mass is determined by the time dependent cosmological expectation value of the Higgs field, for a macroscopic theory one can use Eq.(2) to embed the factor $\phi(t)/\phi_0$ into the metric and the line element becomes
\begin{equation}
ds^2=<\phi(t)/\phi_0>^2 \eta_{\mu\nu} dx^{\mu}dx^{\nu}
\end{equation}
as far as homogeneous and isotropic space-time is considered. Now, while $\phi/\phi_0$ is representing the time dependence of the mass, from another perspective, it can be considered as a part of the metric tensor for a cosmological scenario by the relation
\begin{equation}
a(t)=<\phi(t)/\phi_0>.
\end{equation}
Thus, roughly speaking, the scale factor $a$ must be related to the time dependent cosmological expectation value of the Higgs field.

Besides many successes of the general theory of relativity in explaining some phenomena in the solar system as well as in the standard model of cosmology such as the precession of the mercury, gravitational lensing, proton-neutron ratio in the early universe and the primordial nucleosynthesis, its insufficiency to solve the late-time accelerating expansion of the universe and the galaxy rotation curves without adding dark energy and dark matter as unknown exotic constituents led to search for the modified or the alternative gravity theories\cite{peebles2003}\cite{sahni2004}. It also suffers from some conceptual issues\cite{brans1961}\cite{raine1975} related to Mach's principle which it relies on. Whereas modified gravity theories satisfy WEP, the strong equivalence principle (SEP) is violated as a result of the introduction of a fifth force\cite{brans1961}\cite{joyce2016}. Thus, objects which have different gravitational binding energies, move on different geodesics of the space-time metric. In this sense, the general theory of relativity is the only tensor theory which satisfies both WEP and SEP. Among modified gravity theories, due to the coupling of a simple scalar field to the geometry of space-time, scalar-tensor theories are the more prevalent and flexible alternatives. In this manuscript, we will consider the Jordan Brans Dicke (JBD) theory\cite{brans1961,jordan1959,dicke1959new,dicke1959gravitation} which is the first scalar-tensor theory and seems to be a more complete theory of gravitation with respect to Mach's principle. Furthermore, for our case, it is more suitable to show the relation between the relativistic cosmology and the Higgs mechanism by relating the scalar fields of two picture. In their original paper, Brans and Dicke define the reciprocal of Newton's constant $1/G$ as the scalar field which has the dimension of mass squared. Since our approach will mostly be field theoretical, it is better define a field to have the dimension of mass. Thus, the JBD action extended with a potential term (here, a massless JBD field is taken into account) turns out to be
\begin{equation}\label{eq6}
S=\int d^4x \sqrt{-g}\left(-\frac{\tilde{\xi}^2}{2}\tilde{\chi}^2R-\frac{1}{2}g^{\mu\nu}\partial_\mu \tilde{\chi}\partial_\nu \tilde{\chi}-\frac{\tilde{\lambda}}{4}\tilde{\chi}^4\right).
\end{equation}
Here, $R$, $\tilde{\chi}$ and $\tilde{\xi}^2$ are the Ricci scalar, the JBD scalar field and the dimensionless parameter respectively. As it will be explained at the beginning of the next section, the use of tilde sign in Eq.(6) is because of some dimensional concerns for coordinate transformation in field space. Furthermore, although the coupling parameter $\omega$ which is the original JBD parameter, is more common in the literature, we prefer to stick with the action form in Eq.(6). So, the relation between two parameters is given by $\tilde{\xi}^2=-1/4\omega$.

\section{The Higgs Field and the Conformal Factor in the JBD Theory }
Since it is better fit this scalar-tensor theory into relatively simpler form as the Lagrangian of two scalar field in Minkowski spacetime, we use the following two relations
\begin{equation}
\sqrt{-g}=a^4(t)\sqrt{-\eta}=a^4(t)
\end{equation}
\begin{equation}
R=6\frac{\partial_0^2a}{a^3}
\end{equation}
to write the Lagrangian density in Eq.(6) in terms of dimensionless scalar fields $\chi$ and $a$ as
\begin{equation}\label{lagrangian9}
\mathcal{L}=-\frac{\xi^2}{2}\chi^2 a \partial_0^2a+\frac{\mu^2}{2}a^2(\partial_0\chi)^2-\frac{\lambda}{4}a^4\chi^4 
\end{equation}
where $\chi=\tilde{\chi}/ \mu$ and it has undergone dimensional transmutation. To make $\chi$ dimensionless, $\mu$ must have dimension of mass and so dimensionful constants are defined as $\xi=\mu\tilde{\xi}$ and $\lambda=\mu^4\tilde{\lambda}$. Also, the factor of six in the Ricci scalar has already been embedded within $\xi^2$. 

After the first term in Eq.(9) is expanded by applying integration by parts, the Lagrangian density becomes
\begin{equation}
\mathcal{L}=-\frac{\xi^2}{2}[\partial_0(\chi^2 a \partial_0 a)-2a\chi\partial_0a\partial_0\chi-\chi^2(\partial_0 a)^2]+\frac{\mu^2}{2}a^2(\partial_0\chi)^2-\frac{\lambda}{4}a^4\chi^4 .
\end{equation}
At this point, since the first term in the square bracket is a total divergence, it can be set to zero at the infinity in the action. So, after disregarding this term, by addition and subtraction of the term $\frac{\xi^2}{2}a^2(\partial_0\chi)^2$ into the Lagrangian density, and then taking the factor of $a^2\chi^2$ outside the parenthesis, one ends up with
\begin{equation}
\mathcal{L}=\frac{\xi^2}{2}a^2\chi^2\left[\left(\frac{\partial_0\chi}{\chi}\right)^2+2\frac{\partial_0a}{a}\frac{\partial_0\chi}{\chi}+\left(\frac{\partial_0 a}{a}\right)^2\right]+\frac{\mu^2-\xi^2}{2}a^2(\partial_0\chi)^2-\frac{\lambda}{4}a^4\chi^4 .
\end{equation}
The plus sign in front of the bracket in which the terms correspond the kinetic energy, implies the positivity of $\xi^2$ but negativity of the JBD parameter $\omega$ based upon our definition in the introduction. In order to simplify this expression, the following relation for the terms inside the bracket and the definitions (or the coordinate transformations in the field space) for fields $\alpha$ and $\gamma$ are very useful.
\begin{equation}
[\partial_0(\ln\chi+\ln a)]^2=[\partial_0(\ln\chi a)]^2=[\partial_0\ln\alpha]^2
\end{equation}
\begin{equation}
\alpha=\chi a
\end{equation}
\begin{equation}
\gamma=\ln\chi
\end{equation}
Then, the Lagrangian density can be put in the form
\begin{equation}
\mathcal{L}=\frac{\xi^2}{2}(\partial_0\alpha)^2+\frac{\mu^2-\xi^2}{2}\alpha^2(\partial_0\gamma)^2-\frac{\lambda}{4}\alpha^4.
\end{equation}
Based upon the equation of motion of $\gamma$, since
\begin{equation}
\frac{\partial\mathcal{L}}{\partial\gamma}=0,
\end{equation}
\begin{equation}
\frac{\partial\mathcal{L}}{\partial(\partial_0\gamma)}=(\mu^2-\xi^2)\alpha^2\partial_0\gamma
\end{equation}
must be equal to a constant. Thus,
\begin{equation}
\partial_0\gamma=\frac{C}{(\mu^2-\xi^2)\alpha^2}.
\end{equation}
After some algebra, Hamiltonian density is found as
\begin{equation}
\mathcal{H}=\frac{\xi^2}{2}(\partial_0\alpha)^2+\frac{\mu^2-\xi^2}{2}\alpha^2(\partial_0\gamma)^2+\frac{\lambda}{4}\alpha^4,
\end{equation}
and using Eq.(18) in Eq.(19) yields
\begin{equation}
\mathcal{H}=\frac{\xi^2}{2}(\partial_0\alpha)^2+\frac{C^2}{2(\mu^2-\xi^2)}\frac{1}{\alpha^2}+\frac{\lambda}{4}\alpha^4.
\end{equation}

Before obtaining the equation of motion of $\alpha$, one must have the Hamiltonian density in terms of the field and its canonical momentum. In our case, it will be equal to
\begin{equation}
\mathcal{H}=\frac{1}{2\xi^2}\pi_\alpha^2+\frac{C^2}{2(\mu^2-\xi^2)}\frac{1}{\alpha^2}+\frac{\lambda}{4}\alpha^4,
\end{equation}
where the canonical momentum is
\begin{equation}
\pi_\alpha=\xi^2\partial_0\alpha.
\end{equation}
Furthermore, since the equation of motion is given by
\begin{equation}
-\frac{\partial\mathcal{H}}{\partial\alpha}=\partial_0\pi_\alpha,
\end{equation}
and the left hand side of Eq.(23) is
\begin{equation}
-\frac{\partial\mathcal{H}}{\partial\alpha}=\frac{C^2}{(\mu^2-\xi^2)}\frac{1}{\alpha^3}-\lambda\alpha^3,
\end{equation}
after some algebraic manipulation one can easily get the equation of motion as
\begin{equation}
\partial_0^2\alpha-\frac{C^2}{\xi^2(\mu^2-\xi^2)}\frac{1}{\alpha^3}+\frac{\lambda}{\xi^2}\alpha^3=0.
\end{equation}
In Eq.(21), last two terms behave as an effective potential, so one may write
\begin{equation}
\mathcal{H}=\frac{\xi^2}{2}(\partial_0\alpha)^2+V_{eff},
\end{equation}
where
\begin{equation}
V_{eff}=\frac{C^2}{2(\mu^2-\xi^2)}\frac{1}{\alpha^2}+\frac{\lambda}{4}\alpha^4.
\end{equation}
To determine the vacuum expectation value of the field, the derivative of the potential with respect to $\alpha$ must be equal to zero
\begin{equation}
\frac{\partial V_{eff}}{\partial\alpha}=0.
\end{equation}
This simple procedure gives the vev of $\alpha$
\begin{equation}
\alpha_0=\left(\frac{C^2}{\lambda(\mu^2-\xi^2)}\right)^{1/6}.
\end{equation}
After substituting Eq.(29) in Eq.(18) in order to solve the field $\gamma$ at the vacuum
\begin{equation}
\partial_0\gamma=\frac{C}{(\mu^2-\xi^2)\alpha_0^2}=\left(\frac{\lambda C}{(\mu^2-\xi^2)^2}\right)^{1/3}=D,
\end{equation}
$\gamma$ is found
\begin{equation}
\gamma=\ln\chi=Dt+E,
\end{equation}
where $D$ and $E$ are another constants which must be determined. Then, on the basis of the definition in Eq.(14), the JBD scalar field, is obtained at the vacuum as
\begin{equation}
\chi=e^{Dt+E}.
\end{equation}
Since the temporal evolution of the universe is designated by the scale factor which also gives the time dependence of the Higgs field in our theoretical model, we can take advantage of the definition of $\alpha$ at its vev
\begin{equation}
\alpha_0=a\chi,
\end{equation}
to find
\begin{equation}
a=\frac{\alpha_0}{e^{Dt+E}}=\exp[-D(t-t_0)],
\end{equation}
where
\begin{equation}
\alpha_0 e^{-E}=e^{Dt_0},
\end{equation}
in which $t_0$ is the age of the universe to make the scale factor equal to one today. As it is seen, Eq.(35) implies an exponential expansion for space-time intervals but this is true in comoving time. After one switches to the cosmological time which will be represented with $t^{\prime}$ throughout the manuscript, and arrange constants accordingly to be able to set today's value of $a$ to one, a linear expansion is obtained 
\begin{equation}
a(t^{\prime})=\frac{t^{\prime}}{t_0^{\prime}}.
\end{equation}

We have already learned the evolution of the fields with time at the vev of $\alpha$. Now, a small perturbation can be added to $\alpha$
\begin{equation}
\alpha=\alpha_0(1+\epsilon(t)),
\end{equation}
and insert this into Eq.(25) to get
\begin{equation}
\partial_0^2\epsilon(t)-\frac{C^2}{\xi^2(\mu^2-\xi^2)}\alpha_0^{-4}(1+\epsilon(t))^{-3}+\frac{\lambda}{\xi^2}\alpha_0^2(1+\epsilon(t))^{3}=0.
\end{equation}
Since the perturbation is small in comparison with $\alpha_0$, the second and the third terms in Eq.(38) can be expanded by keeping only the zeroth and the first order terms and it turns out to be
\begin{equation}
\partial_0^2\epsilon(t)-\frac{C^2}{\xi^2(\mu^2-\xi^2)}\alpha_0^{-4}(1-3\epsilon(t))+\frac{\lambda}{\xi^2}\alpha_0^{2}(1+3\epsilon(t))=0.
\end{equation}
Since the zeroth order terms give
\begin{equation}
-\frac{C^2}{\xi^2(\mu^2-\xi^2)}\alpha_0^{-4}+\frac{\lambda}{\xi^2}\alpha_0^{2}=0,
\end{equation}
we are left with the equation to solve
\begin{equation}
\partial_0^2\epsilon(t)+\left(\frac{3C^2}{\xi^2(\mu^2-\xi^2)}\alpha_0^{-4}+\frac{3\lambda}{\xi^2}\alpha_0^{2}\right)\epsilon(t)=0.
\end{equation}
The constant term in the parenthesis has the dimension of mass squared so it may be redefined to write the equation as
\begin{equation}
\partial_0^2\epsilon(t)+m^2\epsilon(t)=0,
\end{equation}
where
\begin{equation}
m^2=\left(\frac{3C^2}{\xi^2(\mu^2-\xi^2)}\alpha_0^{-4}+\frac{3\lambda}{\xi^2}\alpha_0^{2}\right).
\end{equation}
Using the vev of $\alpha$ from Eq.(29) makes $m^2$ to be equal to
\begin{equation}
m^2=\frac{6(\lambda C)^{2/3}}{\xi^2(\mu^2-\xi^2)^{1/3}},
\end{equation}
then the solutions for $\epsilon$ and $\alpha$, around the vacuum, are found as
\begin{equation}
\epsilon(t)=\epsilon_0(e^{imt}+e^{-imt}),
\end{equation}
\begin{equation}
\alpha=\alpha_0(1+\epsilon_0(e^{imt}+e^{-imt})).
\end{equation}
Ultimately, we are interested in the solutions of the fields $\chi$ and $a$. We can follow the same procedure as before by finding $\gamma$ first, then $\chi$ and $a$. To do that Eq.(46) is placed into Eq.(18) again by ignoring second and higher order terms of $\zeta$
\begin{equation}
\partial_0\gamma=\frac{C}{(\mu^2-\xi^2)\alpha^2}=\frac{C}{(\mu^2-\xi^2)}\alpha_0^{-2}\left(1-2\epsilon(t)\right),
\end{equation}
then by using the definition of constant $D$ and integrating 
\begin{equation}
\partial_0\gamma=D(1-2\epsilon(t)),
\end{equation}
$\gamma$ is gained as
\begin{equation}
\gamma=Dt+F+ \frac{i2D}{m}\epsilon_0(e^{imt}-e^{-imt}).
\end{equation}
Here, $F$ is another integration constant which must be defined. Using the relations $\gamma=\ln \chi$ and $\alpha=a\chi$ one more time in order results in
\begin{equation}
\chi(t)=\exp\left(Dt+F+ \frac{i2D}{m}\epsilon_0(e^{imt}-e^{-imt})\right),
\end{equation}
\begin{equation}
a(t)=\alpha_0(1+\epsilon(t)) \exp\left(-Dt-F-\frac{i2D}{m}\epsilon_0(e^{imt}-e^{-imt})\right).
\end{equation}
Once again the higher order terms are disregarded because of the fact that $\epsilon \ll 1$ and constant $F$ is selected to be equal to $E$ in Eq.(31) (since $a(t_0)=1$), so the evolution of the universe in the conformal time is
\begin{equation}
a(t)=\exp\left(-D(t-t_0) + \epsilon_0\left(\left(1- \frac{i2D}{m}\right)e^{imt}+\left(1+ \frac{i2D}{m}\right)e^{-imt}\right)\right),
\end{equation}
and in the cosmological time is
\begin{equation}
a(t^\prime)=\left(\frac{t^\prime}{t_0^\prime}+  \epsilon_0\left(\left(1- \frac{i2D}{m}\right)e^{imt^\prime}+\left(1+ \frac{i2D}{m}\right)e^{-imt^\prime}\right)\right).
\end{equation}
At this point, it is important to note that for the quantization of oscillation modes of $\epsilon$, it can be written in terms of creation and annihilation operators $A$ and $A^\dagger$ like
\begin{equation}
\epsilon(t)=\epsilon_0 (A e^{imt}+ A^{\dagger}  e^{-imt}).
\end{equation}

\section{ Coordinate Transformation in Field Space}

From a non-static cosmological perspective, once the metric is defined as $g_{\mu\nu}=a^2(t) \eta_{\mu\nu}$, one can rewrite the action in Eq.(\ref{eq6}) more explicitly as

\begin{equation}
S=\int d^4x \bigg[\frac{1}{2} \bigg(\xi^2 \chi^2 (\partial_0a)^2 +2\xi^2 a \chi \partial_0 a \partial_0 \chi+ \mu^2 a^2 (\partial_0 \chi)^2\bigg)-\frac{\lambda}{4} a^4\chi^4\bigg].
\end{equation}
 It is easily seen that this action defines a non-linear $\sigma$ model\cite{gell1960}\cite{peskin1995} in which a potential term is added, and can be represented as
\begin{equation} \label{eq55}
S=\frac{1}{2} \int d^4x \left(G_{bc}(\psi)\partial_0 \psi^b \partial_0 \psi^c- V(\psi)\right)
\end{equation}
where $\Psi_{b, c}$ corresponds to $a$ and $\chi$. In addition, the metric in Eq.(\ref{eq55}) is given by

\begin{equation}\label{metric1}
G_{bc}=\begin{pmatrix}
\xi^2 \chi^2 & \xi^2 a \chi\\
\xi^2 a\chi & \mu^2 a^2\\
\end{pmatrix}
\end{equation}
whose scalar curvature can be found zero after straightforward calculations. Since the metric is flat, the kinetic term of this action can be converted to that of Klein-Gordon action by a coordinate transformation in field space. In this way, one can investigate the action in Eq.(55) from the perspective of the Higgs mechanism. Also, when the following transformation between $G_{bc}$ and  $\hat{G}_{bc}$  is achieved, it means we have a non-linear sigma model in the JBD picture.
\begin{equation}
G_{bc}=
\begin{pmatrix}
\xi^2\chi^2& \xi^2 a \chi \\
\xi^2 a \chi& \mu^2 a^2
\end{pmatrix}\longleftrightarrow
\hat{G}_{bc}=
\begin{pmatrix}
1&0\\
0&1
\end{pmatrix}
\end{equation}

Since we are looking for a transformation of the Lagrangian density from the JBD picture to the Higgs picture, the starting point is to write the line element of the target space as
\begin{equation}
ds^2=\frac{1}{2}[\xi^2\chi^2(da)^2+2\xi^2a\chi da d\chi+\mu^2 a^2(d\chi)^2].
\end{equation}
Adding and subtracting the term  $\xi^2 a^2(d\chi)^2$ in Eq.(59), and then taking the factor of $\xi^2a^2\chi^2$ outside the parenthesis results in
\begin{equation}
ds^2=\frac{1}{2}\xi^2 a^2\chi^2\left[\left(\frac{d\chi}{\chi}+\frac{da}{a}\right)^2+\frac{\mu^2-\xi^2}{\xi^2}\left(\frac{d\chi}{\chi}\right)^2\right].
\end{equation}
Relating $a$ and $\chi$ to new fields $\alpha$ and $\gamma$ as we did before in Eq.(13) and Eq.(14), gives
\begin{equation}
a(\alpha, \gamma)=\alpha e^{-\gamma},
\end{equation}
\begin{equation}
\frac{da}{a}=\frac{d\alpha}{\alpha}-d\gamma,
\end{equation}
\begin{equation}
\chi(\gamma)=e^{\gamma},
\end{equation}
\begin{equation}
\frac{d\chi}{\chi}=d\gamma,
\end{equation}
the line element in Eq.(60) turns out to be
\begin{equation}
ds^2=\frac{1}{2}\xi^2\left(d\alpha^2+\frac{\mu^2-\xi^2}{\xi^2}\alpha^2 d\gamma^2\right).
\end{equation}
At this point, another transformation is needed to get rid of the factors and the following ones are useful to accomplish this. 
\begin{equation}
\alpha(\rho)=\frac{\rho}{\xi}
\end{equation}
\begin{equation}
d\alpha=\frac{d\rho}{\xi}
\end{equation}
\begin{equation}
\gamma(\theta)=\frac{\xi}{\sqrt{\mu^2-\xi^2}}\theta
\end{equation}
\begin{equation}
d\gamma=\frac{\xi}{\sqrt{\mu^2-\xi^2}} d\theta
\end{equation}
Substitution of Eq.(67) and Eq.(69) into Eq.(65) yields
\begin{equation}
ds^2=\frac{1}{2}d\rho^2+\frac{1}{2}\rho^2 d\theta^2.
\end{equation}
Here, $\rho$ and $\theta$ correspond to spherical coordinates. To get $\hat{G}_{\mu\nu}$ in Eq.(59), it is straightforward to define them as
\begin{equation}
\rho(\phi_3, \phi_5)=\sqrt{\phi_3^2+\phi_5^2},
\end{equation}
\begin{equation}
\theta(\phi_3, \phi_5)=\arctan\frac{\phi_5}{\phi_3}.
\end{equation}
When these new coordinates are used in Eq.(70), one can write the line element as desired from the very beginning of this section and it is
\begin{equation}
ds^2=\frac{1}{2}(d\phi_3^2+d\phi_5^2).
\end{equation}
To write the coordinates $a$ and $\chi$ in terms of $\phi_3$ and $\phi_5$, all the transformations can be applied one by one from the beginning to the end. First of all, after implementing Eq.(66) and Eq.(68) into Eq.(61) and Eq.(63), $a$ and $\chi$ can be expressed like
\begin{equation}
a(\rho, \theta)=\frac{\rho}{\xi}\exp\left(-\frac{\xi}{\sqrt{\mu^2-\xi^2}}\theta\right),
\end{equation}
\begin{equation}
\chi(\theta)=\exp\left(\frac{\xi}{\sqrt{\mu^2-\xi^2}}\theta\right).
\end{equation}
Then, the transformation from the spherical coordinates to the cartesian ones results in
\begin{equation}
a(\phi_3, \phi_5)=\frac{\sqrt{\phi_3^2+\phi_5^2}}{\xi}\exp\left(-\frac{\xi}{\sqrt{\mu^2-\xi^2}}\arctan\left(\frac{\phi_5}{\phi_3}\right)\right),
\end{equation}
\begin{equation}
\chi(\phi_3, \phi_5)=\exp\left(\frac{\xi}{\sqrt{\mu^2-\xi^2}}\arctan\left(\frac{\phi_5}{\phi_3}\right)\right).
\end{equation}

At this point, it is also possible to state $\phi_3$ and $\phi_5$ in terms of $a$ and $\chi$ by carrying out all the transformations back in order. To start with, because of the spherical ones which have lastly been obtained, $\phi_3$ and $\phi_5$ are
\begin{equation}
\phi_3(\rho, \theta)=\rho\cos\theta,
\end{equation}
\begin{equation}
\phi_5(\rho, \theta)=\rho\sin\theta.
\end{equation}
Thanks to Eq.(66) and Eq.(68), $\rho$ and $\theta$ are found
\begin{equation}
\rho(\alpha)=\xi\alpha,
\end{equation}
\begin{equation}
\theta(\gamma)=\frac{\sqrt{\mu^2-\xi^2}}{\xi}\gamma,
\end{equation}
and then using Eq.(80) and Eq.(81) in Eq.(78) and Eq.(79) gives
\begin{equation}
\phi_3(\alpha, \gamma)=\xi\alpha\cos \left(\frac{\sqrt{\mu^2-\xi^2}}{\xi}\gamma \right),
\end{equation}
\begin{equation}
\phi_5(\alpha, \gamma)=\xi\alpha\sin \left(\frac{\sqrt{\mu^2-\xi^2}}{\xi}\gamma \right).
\end{equation}
Since, on the basis of Eq.(61) and Eq.(63), $\alpha$ and $\gamma$ are
\begin{equation}
\alpha(a, \chi)=a\chi,
\end{equation}
\begin{equation}
\gamma(\chi)=\ln\chi,
\end{equation}
substituting these into Eq.(82) and Eq.(83) gives the scalar fields of the Higgs picture $\phi_3$ and $\phi_5$ in terms of those of the JBD picture $a$ and $\chi$ as
\begin{equation}\label{scalarfactor1}
\phi_3(a, \chi)=\xi a \chi\cos\left(\frac{\sqrt{\mu^2-\xi^2}}{\xi}\ln\chi\right),
\end{equation}
\begin{equation}\label{scalarfactor2}
\phi_5(a, \chi)=\xi a \chi\sin\left(\frac{\sqrt{\mu^2-\xi^2}}{\xi}\ln\chi\right).
\end{equation}

Therefore, in terms of  $\phi_3$ and $\phi_5$, the Lagrangian density in Eq.(55) can be stated as 

\begin{equation} \label{simplelag}
\mathcal{L} =\frac{1}{2} (\partial_0\phi_3)^2 +\frac{1}{2} (\partial_0\phi_5)^2-\frac{\kappa}{4} (\phi^2_3+\phi^2_5)^2
\end{equation}
where $\kappa=\lambda \xi^{-4}$.

\section{The Higgs Picture}

The Lagrangian density of the Higgs field in doublet form is taken to be

\begin{equation}
\mathcal{L}= \partial_\mu \Theta^\dagger \partial^\mu \Theta -V(\Theta)
\end{equation}
with $\Theta=\frac{1}{\sqrt{2}}\begin{pmatrix}
\phi_1+i\phi_2\\
\phi_3+i\phi_4 
\end{pmatrix}$
where $\phi_a$ corresponds to scalar fields and  $a=1,2,3,4$. Furthermore, the potential term can be defined as

\begin{equation}
V(\Theta)=-\frac{1}{2}\bar{m}^2\Theta^\dagger\Theta+\frac{\kappa}{4}(\Theta^\dagger\Theta)^2
\end{equation}
where dimensionless constant $\kappa  \textgreater  0$ and the scalar fields have dimension of mass. In addition, the mass term has a minus sign so that for a time-independent expectation value, spontaneous symmetry breaking occurs. In terms of the fields $\phi_a$, the Lagrangian density can be written as

\begin{equation} \label{smpot} 
\mathcal{L} =\frac{1}{2} \partial_\mu\phi_a  \partial^\mu\phi^a-\frac{\kappa}{4} (\phi_a\phi^a)^2
\end{equation} 
where we put $\bar{m}=0$ so that the potential term in Eq.(\ref{smpot}) is purely quartic.

Note that the symmetry of this Lagrange density is SO(5) which is larger than the gauge symmetry  $SU(2)\times U(1)$ of the standard model. We will extend this Lagrangian by adding an additional scalar field $\phi_5 $, so that now

$$a=1,2,3,4,5.$$

Since the rotational symmetry  is spontaneously broken, a  fluctuation emerges about the minimum. Breaking the symmetry annihilates three of four  components of $\Theta$ such that

\begin{equation}  
\phi_1=\phi_2=\phi_4=0.
\end{equation}
Moreover, the fields can be independent of spatial coordinates to be transformable to those of the Jordan-Brans-Dicke theory. Then one obtains the Lagrangian density in Eq.(\ref{simplelag}). 

At this point, applying field space coordinate transformations in Eq.(\ref{scalarfactor1}) and Eq.(\ref{scalarfactor2}) to the vacuum expectation values and their quantum fluctuations of the fields of the JBD theory in order to find their correspondence in the Higgs mechanism gives
\begin{equation}  \label{dim1}
\begin{split}
\phi_3=\xi \chi_0  \cos(H(t)) \bigg(&1+\epsilon_0\big( A e^{imt}+  A^\dagger e^{-imt}\big)\\
&-i\sqrt{\frac{2}{3}}\tan(H(t)) \epsilon_0  \big( A e^{imt}-  A^\dagger e^{-imt}\big)\bigg)
\end{split}
\end{equation}

\begin{equation}  \label{dim2}
\begin{split}
\phi_5=\xi \chi_0  \sin(H(t)) \bigg(&1+\epsilon_0\big( A e^{imt}+  A^\dagger e^{-imt}\big)\\
&+i\sqrt{\frac{2}{3}}\cot(H(t)) \epsilon_0  \big( A e^{imt}-  A^\dagger e^{-imt}\big)\bigg)
\end{split}
\end{equation}
where
\begin{equation}
H(t)=\frac{\sqrt{\mu^2-\xi^2}}{\xi} (Dt+E),
\end{equation}
and
\begin{equation}
\chi_0=e^{Dt_0+E}
\end{equation}
which is the today's value of the JBD field.

We note that the system can be quantized by imposing the commutation relation

\begin{equation}\label{eqqu1}
\left[ A,A^\dagger\right]=1.
\end{equation}
Here, $A$ and $A^\dagger$ are the creation and annihilation operators of the quantum particles and the vucuum expectation values of $\phi_3$ and $\phi_5$ are given by

\begin{equation} \label{vav1}
\langle\phi_3 \rangle=\xi \chi_0  \cos\left(\frac{\sqrt{\mu^2-\xi^2}}{\xi}Dt\right),
\end{equation} 
 
\begin{equation} \label{vav2}
\langle\phi_5 \rangle= \xi \chi_0  \sin\left(\frac{\sqrt{\mu^2-\xi^2}}{\xi}Dt\right).
\end{equation}  

Here, the temporal evolution of the vev of the Higgs field is given by the argument of cosine in Eq.(98), i.e. the parameter $D$. As it can be checked by relating the scale factors in two different time scales (conformal and cosmological time) in Eq.(34) and Eq.(36), $D=-\frac{1}{t^\prime_0}$ in which $t^\prime$ is the age of the universe in cosmological time. Thus, in our model $D$ and the evolution of the Higgs field are very slow and the condition about the particle masses, which has been mentioned before Eq.(2) in the introduction, is satisfied.

\section{Conclusion}

A cosmological model in which the expansion of the universe is related to the time dependent vev of the Higgs field has been proposed. Based upon Eq.(1), the time dependent inertial mass may have another interpretation such that the time dependence of the Higgs field is part of the metric tensor. With this approach, the Higgs field has been taken into account as a conformal factor and related to the scale factor of the FLRW metric. Since it is a more complete theory of gravitation with respect to Mach's principle, the JBD theory has been considered and only the scalar mode of the theory has been studied. By taking the action of the scale factor $a(t)$ and the JBD field $\chi(t)$ as depending only on time, the relation between the JBD cosmology and the Higgs mechanism has been established with the field space coordinate transformations (Eq.(76), Eq.(77), Eq.(86) and Eq.(87)) for negative values of the JBD parameter. Although solar system experiments predict the original JBD parameter $\omega$ to be a big positive number\cite{li2013}\cite{will2018}, scenarios based on its negative values\cite{bertolami2000,sen2001,sen2003,banerjee2001,batista2001,fabris2005} are viable and quite common in the literature for cosmological scales. In addition to this, negative values of the coupling parameter are encountered in the applications of the low-energy effective action of the string theory\cite{fabris2003regular}\cite{calcagni2005} such that the dilatonic coupling constant is chosen as $\omega=-1$ for the string frame\cite{copeland1994,gasperini1996,wands2002}. Finally, oscillation modes about the vacuum in both pictures have been found and it has been shown that they are quantizable.

\nocite{*}
\bibliographystyle{ieeetr}
\bibliography{references}

\begin{thebibliography}{10}

\bibitem{2012observation}
G.~Aad, T.~Abajyan, B.~Abbott, J.~Abdallah, S.~A. Khalek, A.~A. Abdelalim,
  R.~Aben, B.~Abi, M.~Abolins, O.~AbouZeid, {\em et~al.}, ``Observation of a
  new particle in the search for the standard model higgs boson with the atlas
  detector at the lhc,'' {\em Physics Letters B}, vol.~716, no.~1, pp.~1--29,
  2012.

\bibitem{higgs1964}
P.~W. Higgs, ``Broken symmetries and the masses of gauge bosons,'' {\em
  Physical Review Letters}, vol.~13, no.~16, p.~508, 1964.

\bibitem{weinberg1991}
S.~Weinberg, ``Conceptual foundations of the unified theory of weak and
  electromagnetic interaction,'' in {\em Origin Of Symmetries}, pp.~215--223,
  World Scientific, 1991.

\bibitem{salam1980}
A.~Salam, ``Gauge unification of fundamental forces,'' {\em Reviews of Modern
  Physics}, vol.~52, no.~3, p.~525, 1980.

\bibitem{glashow1980}
S.~L. Glashow, ``Towards a unified theory: Threads in a tapestry,'' {\em
  Reviews of Modern Physics}, vol.~52, no.~3, p.~539, 1980.

\bibitem{eotvos1922}
R.~v. Eotvos, ``Beitrage zum gesetze der proportionalitat von tragheit und
  gravitat,'' {\em Ann. Phys.}, vol.~68, pp.~11--66, 1922.

\bibitem{sciama1953}
D.~W. Sciama, ``On the origin of inertia,'' {\em Monthly Notices of the Royal
  Astronomical Society}, vol.~113, no.~1, pp.~34--42, 1953.

\bibitem{arik2020}
M.~Arik and T.~Tok, ``The scalar mode of gravity,'' {\em arXiv preprint
  arXiv:2001.02347}, 2020.

\bibitem{friedmann1922}
A.~Friedman, ``{\"U}ber die kr{\"u}mmung des raumes,'' {\em Zeitschrift f{\"u}r
  Physik}, vol.~10, no.~1, pp.~377--386, 1922.

\bibitem{lemaitre1931}
G.~Lema{\^\i}tre, ``Expansion of the universe, the expanding universe,'' {\em
  Monthly Notices of the Royal Astronomical Society}, vol.~91, pp.~490--501,
  1931.

\bibitem{robertson1935}
H.~P. Robertson, ``Kinematics and world-structure,'' {\em The Astrophysical
  Journal}, vol.~82, p.~284, 1935.

\bibitem{walker1937}
A.~G. Walker, ``On milne's theory of world-structure,'' {\em Proceedings of the
  London Mathematical Society}, vol.~2, no.~1, pp.~90--127, 1937.

\bibitem{peebles2003}
P.~J.~E. Peebles and B.~Ratra, ``The cosmological constant and dark energy,''
  {\em Reviews of modern physics}, vol.~75, no.~2, p.~559, 2003.

\bibitem{sahni2004}
V.~Sahni, ``Dark matter and dark energy,'' in {\em The Physics of the Early
  Universe}, pp.~141--179, Springer, 2004.

\bibitem{brans1961}
C.~Brans and R.~H. Dicke, ``Mach's principle and a relativistic theory of
  gravitation,'' {\em Physical review}, vol.~124, no.~3, p.~925, 1961.

\bibitem{raine1975}
D.~J. Raine, ``Mach's principle in general relativity,'' {\em Monthly Notices
  of the Royal Astronomical Society}, vol.~171, no.~3, pp.~507--528, 1975.

\bibitem{joyce2016}
A.~Joyce, L.~Lombriser, and F.~Schmidt, ``Dark energy versus modified
  gravity,'' {\em Annual Review of Nuclear and Particle Science}, vol.~66,
  pp.~95--122, 2016.

\bibitem{jordan1959}
P.~Jordan, ``Zum gegenw{\"a}rtigen stand der diracschen kosmologischen
  hypothesen,'' {\em Zeitschrift f{\"u}r Physik}, vol.~157, no.~1,
  pp.~112--121, 1959.

\bibitem{dicke1959new}
R.~H. Dicke, ``New research on old gravitation,'' {\em Science}, vol.~129,
  no.~3349, pp.~621--624, 1959.

\bibitem{dicke1959gravitation}
R.~H. Dicke, ``Gravitation—an enigma,'' {\em American Scientist}, vol.~47,
  no.~1, pp.~25--40, 1959.

\bibitem{gell1960}
M.~Gell-Mann and M.~L{\'e}vy, ``The axial vector current in beta decay,'' {\em
  Il Nuovo Cimento (1955-1965)}, vol.~16, no.~4, pp.~705--726, 1960.

\bibitem{peskin1995}
M.~E. Peskin and D.~V. Schroeder, ``An introduction to quantum field theory
  (boulder, co),'' 1995.

\bibitem{li2013}
Y.-C. Li, F.-Q. Wu, and X.~Chen, ``Constraints on the brans-dicke gravity
  theory with the planck data,'' {\em Physical Review D}, vol.~88, no.~8,
  p.~084053, 2013.

\bibitem{will2018}
C.~M. Will, {\em Theory and experiment in gravitational physics}.
\newblock Cambridge university press, 2018.

\bibitem{bertolami2000}
O.~Bertolami and P.~Martins, ``Nonminimal coupling and quintessence,'' {\em
  Physical Review D}, vol.~61, no.~6, p.~064007, 2000.

\bibitem{sen2001}
S.~Sen and A.~Sen, ``Late time acceleration in brans-dicke cosmology,'' {\em
  Physical Review D}, vol.~63, no.~12, p.~124006, 2001.

\bibitem{sen2003}
S.~Sen and T.~Seshadri, ``Self interacting brans--dicke cosmology and
  quintessence,'' {\em International Journal of Modern Physics D}, vol.~12,
  no.~03, pp.~445--460, 2003.

\bibitem{banerjee2001}
N.~Banerjee and D.~Pavon, ``A quintessence scalar field in brans-dicke
  theory,'' {\em Classical and Quantum Gravity}, vol.~18, no.~4, p.~593, 2001.

\bibitem{batista2001}
A.~Batista, J.~Fabris, and R.~de~Sa~Ribeiro, ``A remark on brans--dicke
  cosmological dust solutions with negative $\omega$,'' {\em General Relativity
  and Gravitation}, vol.~33, no.~7, pp.~1237--1244, 2001.

\bibitem{fabris2005}
J.~Fabris, S.~Gon{\c{c}}alves, and R.~Ribeiro, ``Late time accelerated
  brans-dicke pressureless solutions and the supernovae type ia data,'' {\em
  arXiv preprint astro-ph/0510779}, 2005.

\bibitem{fabris2003regular}
J.~Fabris, R.~Furtado, P.~Peter, and N.~Pinto-Neto, ``Regular cosmological
  bouncing solutions in low energy effective action from string theories,''
  {\em Physical Review D}, vol.~67, no.~12, p.~124003, 2003.

\bibitem{calcagni2005}
G.~Calcagni, S.~Tsujikawa, and M.~Sami, ``Dark energy and cosmological
  solutions in second-order string gravity,'' {\em Classical and Quantum
  Gravity}, vol.~22, no.~19, p.~3977, 2005.

\bibitem{copeland1994}
E.~J. Copeland, A.~Lahiri, and D.~Wands, ``Low energy effective string
  cosmology,'' {\em Physical Review D}, vol.~50, no.~8, p.~4868, 1994.

\bibitem{gasperini1996}
M.~Gasperini, J.~Maharana, and G.~Veneziano, ``Graceful exit in quantum string
  cosmology,'' {\em Nuclear Physics B}, vol.~472, no.~1-2, pp.~349--360, 1996.

\bibitem{wands2002}
D.~Wands, ``String-inspired cosmology,'' {\em Classical and Quantum Gravity},
  vol.~19, no.~13, p.~3403, 2002.

\end{thebibliography}

\end{document}